\documentclass[fleqn,10pt]{wlscirep}
\title{Optimization of neural networks via finite-value quantum fluctuations}

\author[1,*]{Masayuki Ohzeki}
\author[2]{Shuntaro Okada}
\author[2]{Masayoshi Terabe}
\author[2]{Shinichiro Taguchi}
\affil[1]{Graduate School of Information Sciences, Tohoku University, Sendai 
980-8579, Japan}
\affil[2]{DENSO Corporation, Kariya, Aichi 474-0025, Japan}

\affil[*]{mohzeki@tohoku.ac.jp}

\begin{abstract}
We numerically test an optimization method for deep neural networks (DNNs) using quantum fluctuations inspired by quantum annealing. For efficient optimization, our method utilizes the quantum tunneling effect beyond the potential barriers.
The path integral formulation of the DNN optimization generates an attracting force to simulate the quantum tunneling effect.
In the standard quantum annealing method, the quantum fluctuations will vanish at the last stage of optimization.
In this study, we propose a learning protocol that utilizes a finite value for quantum fluctuations strength to obtain higher generalization performance, which is a type of robustness.
We demonstrate the performance of our method using two well-known open datasets: the MNIST dataset and the Olivetti face dataset.
Although computational costs prevent us from testing our method on large datasets with high-dimensional data, results show that our method can enhance generalization performance by induction of the finite value for quantum fluctuations.
\end{abstract}
\begin{document}

\flushbottom
\maketitle
% * <john.hammersley@gmail.com> 2015-02-09T12:07:31.197Z:
%
%  Click the title above to edit the author information and abstract
%
\thispagestyle{empty}

\section*{Introduction}
Data-driven approach is being widely adopted in many science and engineering fields.
The key technology is machine learning, which is supported by successful examples of the use of deep neural networks (DNNs) \cite{LeCun2015}.
Deep neural networks have achieved state-of-the-art results in a wide variety of tasks, including computer vision, natural language processing, and reinforcement learning \cite{Bengio2016}.
The revolutionary event in which artificial intelligence bested a human at a game of Go exemplifies the potential power of machine learning.
In DNNs, iterative structures of linear and non-linear transformations construct a pattern-recognition system for designing a feature
extractor from the raw data (such as the pixel values of natural image data) into a nontrivial internal representation or feature vector.
The extracted features enable us to classify the different patterns from the input data.

To promote DNN technology, various researchers have developed learning algorithms to provide faster results and better performance.
The algorithms for optimizing DNNs are based on the stochastic gradient descent \cite{Robbins1951,Bottou1998,Sutskever2013}; it partitions a large dataset into several batches and approximates the gradient of the cost function.
The standard choice among the various algorithms stemming from the stochastic gradient method is the Adaptive Momentum (Adam) algorithm \cite{Adam2015}. 
This algorithm is designed to efficiently escape saddle points that often appear in the cost functions of DNNs.
In practice, however, the learning of DNNs suffers from local minima with different generalization performance resulting from the shape of the DNN cost functions.
The sharp minimizer has poorer generalization performance than that in the wide-flat minimizer.
It is thus important to design a learning algorithm to find a more optimal solution by escaping from both the saddle points and the local minima.
In a recent study \cite{SBLB2016}, the batch size is closely related to the generalization performance, which is characterized by the shape of the local minima.
They experimentally demonstrate that the large-batch stochastic gradient method and its variants tend to converge to sharp minimizers with poor generalization performance.
The small-batch stochastic gradient descent, on the other hand, is likely to fall into the wider minimizers, in which the DNNs have high generalization performance.
The batch size is closely related to the magnitude of the stochastic noise during learning.
In other words, injection of the stochastic noise can be an origin of an efficient learning algorithm for converging into wider local minima.
In addition, an analytical study on discrete-weight networks revealed the subdominant solutions with relatively higher generalization performance than the exponentially dominant (typical) solutions that deviated from the ground truth \cite{Baldassi2015, Baldassi2016}.
The subdominant solutions can be algorithmically reachable by considering the effect of entropy.
As proposed in the literature \cite{ESGD}, they compute the local entropy by injection of stochastic noise and update the weight to take the DNN to wider local minima with better generalization performance.

The gradient descent algorithm is closely related to classical dynamics in physics, and the stochastic version also has a connection with Langevin dynamics, which models the classical stochastic dynamics in various fields of nature.
In the present study, we test the optimization of DNNs using the quantum fluctuation as employed in quantum annealing (QA).
Quantum annealing is a method that is developing as a generic solver for the optimization problems.
This scheme was originally proposed as an algorithm that used numerical computations to optimize cost functions with discrete variables  \cite{Kadowaki1998}.
The theoretical aspects of QA are well known. Its basic concept is derived from the quantum adiabatic theorem \cite{Suzuki2005,Morita2008,Ohzeki2011c}, and a successful experimental implementation of QA was realized using present-day technology \cite{Dwave2010a,Dwave2010b,Dwave2010c,Dwave2014}. 
Since then, QA has been developed rapidly and has attracted much attention.
Several protocols based on QA do not stick to the adiabatic quantum computation or maintain the system at the ground state; rather, they employ a nonadiabatic counterpart \cite{Ohzeki2010a,Ohzeki2011,Ohzeki2011proc,Somma2012}.
In addition, some studies have used a more sophisticated quantum effect \cite{Seki2012,Nishimori2017,Ohzeki2017}.
Although the original proposal for QA was designed for optimization problems with discrete variables, as described in the form of a spin-glass Hamiltonian \cite{Kadowaki1998}, the concept of QA can be generalized to a wider range of optimization problems, even those with continuous values.
Most practical optimization problems, including machine learning, use continuous variables.
One typical instance is the optimization problem for DNNs.
Below, we apply the concept of QA to the DNN optimization problem.
In the previous study, they assessed the potential efficiency of using quantum fluctuations to avoid the non-convex cost function by means of the replica method, which is a sophisticated tool in statistical mechanics\cite{Baldassi2018}.
Although the analysis in the previous study discussed the learning of the discrete-weight neural network (binary variable as in the Ising model), the essential features are expected not to differ from the continuous-variable neural networks.
As discussed in the previous study, the generalization performance attained by the optimization with quantum fluctuations can be better than that without them.
In the present study, we perform practical tests: the optimization of DNNs with quantum fluctuations, and discuss its efficiency.
Because the computational cost for simulating quantum dynamics is prohibitive, as shown below, our test is restricted to the case for the relatively shallow networks.
However our approach is straightforward to apply deeper networks.

The paper is organized as follows:
The second section describes our method for optimizing DNNs.
The following section demonstrates the method using three simple tasks.
The last section discusses the feasibility of our method.
\section*{Methods}

\subsection*{Quantum annealing for continuous variables}
The optimization problem is interpreted as the minimization of the energy function (potential energy) $V({\bf w})$ in the context of physics.
We address the optimization of the weights of DNNs below.
The weights are denoted by ${\bf w} \in \mathbb{R}^N$.
The standard gradient descent is given as the equation of motion for the overdamped system
\begin{equation}
{\bf w}(t+1) = {\bf w}(t) - \eta \frac{\partial}{\partial {\bf w}} V({\bf w}).\label{SGD}
\end{equation}
where $t$ is the update step.
This is regarded as a dynamical system in a low-temperature region in the context of physics.
Considering the thermal effect characterized by the temperature $T$, the weights fluctuate following the Gibbs-Boltzmann distribution as
\begin{equation}
P({\bf w}) = \frac{1}{Z}\exp\left( - \beta V({\bf w})\right),
\end{equation}
where $Z$ is the partition function that acts as a normalization constant.
In this case, instead of the equation of motion, a dynamical system with Langevin dynamics is adequate for description of the weights following the Gibbs--Boltzmann distribution as 
\begin{equation}
{\bf w}(t+1) = {\bf w}(t) - \eta \frac{\partial}{\partial {\bf w}} V({\bf w}) + \sqrt{2T\eta}N(0,1).
\end{equation}
This is the procedure known as the stochastic gradient Langevin method \cite{SGLD}, in which the learning rate decreases in the same manner as in simulated annealing (SA) \cite{Kirkpatrick1983}.
In QA, we introduce quantum fluctuations in addition to the energy function in the extremely low temperature $T \to 0 (\beta \to \infty)$.
We consider the following time-dependent Hamiltonian:
\begin{equation}
\hat{H}(t) = V(\hat{{\bf w}}) + \frac{1}{2\rho(t)}\hat{{\bf p}}^2
\end{equation}
where $\hat{{\bf w}}$ denotes degrees of freedom and $\hat{{\bf p}}$ represents momentum that satisfies the commutation relation $[\hat{{\bf w }},\hat{{\bf p}}] = i \hbar.$
In addition, $\rho(t)$ represents the mass of the weights and increases from $0$ to $\infty$ over time throughout the QA process.
Following the ideas of quantum mechanics, the weights fluctuate as characterized by the following density matrix, instead of directly by the distribution function; this is defined as
\begin{equation}
\hat{\rho} = \frac{1}{Z}\exp\left( -\beta \hat{H}(t)\right)
\end{equation}
where $Z = {\rm Tr}\left(\exp(-\beta \hat{H}(t))\right)$.
To specify the probability distribution of the realized configuration of the weights, we compute the matrix elements as
\begin{equation}
P({\bf w}) = \left\langle {\bf w} \right| \hat{\rho} \left|{\bf w}\right\rangle.\label{qdis}
\end{equation}
where $\hat{{\bf w}}\left|{\bf w}\right\rangle = {\bf w} \left|{\bf w}\right\rangle$.
However, the computation of the density matrix is intractable in general.
We then employ the Suzuki--Trotter decomposition to reduce the operators to c-numbers by introducing $M$ copies \cite{Hatano1998} and obtain the following path-integral representation as shown in Appendix:
\begin{equation}
P({\bf w}) = \lim_{M \to \infty} \int \mathcal{D}{\bf w} \exp\left( - \frac{\beta}{M} V({\bf w}_k) - \frac{M \rho(t)}{2\beta} \left\| {\bf w}_{k} - {\bf w}_{k-1}\right\|_2^2 \right).\label{EqSuzuki}
\end{equation}
where $\int \mathcal{D}{\bf w} = \prod_{k=1}^{M-1} \int d{\bf w}_k$, $M$ is the Trotter number and $k$ is the index of the replicated system.
The boundary condition is set to ${\bf w}_0 = {\bf w}_{M} = {\bf w}$.
The numerical implementation of the Suzuki-Trotter decomposition is established as an approximation of the distribution function (\ref{EqSuzuki}) by setting a finite number for $M$.
For instance, in the quantum Monte Carlo simulation \cite{Suzuki1976}, the configuration of the degrees of freedom is sampled using the distribution function as
\begin{equation}
P({\bf w}_1,{\bf w}_2,\cdots,{\bf w}_M) = \prod_{k=1}^{M} \exp\left( - \frac{\beta}{M} V({\bf w}_k) - \frac{M\rho(t)}{2\beta} \left\| {\bf w}_{k} - {\bf w}_{k-1}\right\|_2^2 \right),\label{cost}
\end{equation}
in which the inverse temperature is taken to be $\beta \to \infty$ with $\beta/M$ being finite.
In other words, the quantum Monte Carlo simulation deals with many replicated realizations or paths ${\bf w}_k(t)$ with index $k$ (imaginary time) following Langevin dynamics as 
\begin{equation}
{\bf w}_k(t+1) = {\bf w}_k(t) - \eta \frac{\partial}{\partial {\bf w}_k} V({\bf w}_k(t)) - \eta T_q^2 \rho(t)\left(2{\bf w}_k(t) - {\bf w}_{k-1}(t)- {\bf w}_{k+1}(t)\right) + \sqrt{2T_q\eta}N(0,1).\label{QAdeep}
\end{equation}
where $T_q = M/\beta$.
One might recognize that many DNN realizations interact with each other through the elastic term, which represents the quantum effect.
The elastic term urges many DNN realizations into a single condensed solution ${\bf w}^*$ when $\rho(t)$ takes relatively a large value.
By the boundary condition ${\bf w}_0 = {\bf w}_M$, ${\bf w}^* = {\bf w}$.
For simplicity, let us first consider the case with a large $\rho(t)$.
The path integral formulation allows fluctuation around ${\bf w}^*$.
In other words, the action in the exponential function in $P({\bf w})$ has two terms: one is the cost function, which is what we originally want to optimize, and the other is degree of condensation of the realizations. 
As in Appendix, we find that ${\bf w}_k - {\bf w}$ follows a Gaussian distribution with some covariance $\beta V_{kk'}(t)$.
Thus, the approximated distribution function in a large $\rho(t)$ is reduced to 
\begin{equation}
P({\bf w}) \approx \int \mathcal{D}{\bf w}_k \exp\left( -\frac{\beta}{M}\sum_{k} V({\bf w}_k)\right)\exp\left( - \frac{\beta}{2}\sum_{k,k'}({\bf w}_k - {\bf w})V_{kk'}(t)({\bf w}_{k' }- {\bf w})\right) 
\label{dis}
\end{equation}
Here, we set the minimizer of the (logarithm of) the distribution function in order to make analysis simpler.
\begin{equation}
\log P({\bf w}) \ge M \log \int d{\bf w}' \exp\left( - \frac{\gamma(t)}{2T_q}({\bf w}' - {\bf w})^2 -\frac{1}{T_q}V({\bf w}')\right)\label{MM}
\end{equation}
where $M \gamma$ is a constant for maintaining this inequality.
The minimizer on the right-hand side is the cost function appearing in the entropy stochastic gradient descent (E-SGD) algorithm, which captures the wider local minima \cite{Baldassi2016}.
In order to obtain the most probable weights ${\bf w}$, taking the derivative with respect to ${\bf w}$ of the minimizer of $\log P({\bf w})$, we obtain the following update equation
\begin{equation}
{\bf w}(t) = \gamma(t) \left({\bf w}(t) - \left\langle {\bf w}'\right\rangle\right),
\end{equation}
where $\langle \cdots \rangle$ takes the average of ${\bf w}'$ in the integrand of (\ref{MM}).
The average is directly intractable and is instead estimated by the following Langevin dynamics:
\begin{equation}
{\bf w}'(s+1) = {\bf w}'(s) - \eta \left\{ \frac{\partial}{\partial {\bf w}} V({\bf w}) + \gamma(t) ({\bf w}(t)- {\bf w}'(s))\right\} +  \sqrt{2T_q \eta}N(0,1).
\end{equation}
In the E-SGD algorithm, $\gamma(t)$ is a decreasing value, which will vanish at the completion of optimization.
The time dependence of $\gamma(t)$ is closely related to $\rho(t)$ as described in the Appendix.
In standard QA, we gradually increase $\rho(t)$.
Then $\gamma(t)$ similarly increases.
Thus, the E-SGD algorithm is essentially different from the standard QA procedure.
As they stated, the ``reverse annealing" method is considered in the literature \cite{Baldassi2016}.

%conclusionへ?
Reverse annealing is now implemented in the current system of the D-Wave machine, and shows better performance for optimization.
A similar approach for increasing the performance is to search by induction of quantum fluctuation \cite{Perdomo2012}.
In these cases, reverse annealing is induction of the quantum fluctuation, namely $\rho(0)=\rho(T) = 0$ while $\rho(t)>0$.

\subsection*{Finite-value quantum annealing}
As described in previous studies \cite{Baldassi2016,Baldassi2018}, there is a useful algorithm exploiting an entropic effect around a single condensed solution.
In this algorithm, the author can elucidate one of the aspects related to the quantum effect: i.e., the entropy effect.
In our study, we perform the direct optimization of the cost function, which appears in the exponential of the probability distribution (\ref{cost}), which involves nontrivial quantum tunneling stemming from non-perturbative effects.
Thus, we must deal with $M$ replicated systems for optimizing the DNNs.
In this sense, our procedure is not reasonable for optimizing DNNs in practical applications. 
However, our trial may stimulate motivation for possible applications of the quantum computation.
We report several simple DNN optimization tests to provide future perspectives in machine learning with respect to the quantum mechanics described below.

From this point forward, we do not focus on cases with a large $\rho(t)$.
We consider directly optimizing the cost function (\ref{cost}), but $T \to 0$ in order to obtain only the quantum effect for simplicity, as
\begin{equation}
{\bf w}_k(t+1) = {\bf w}_k(t) - \eta \frac{\partial}{\partial {\bf w}_k} V({\bf w}_k(t)) - \eta \rho(t) \left(2{\bf w}_k(t) - {\bf w}_{k-1}(t)- {\bf w}_{k+1}(t)\right).\label{RQA}
\end{equation}
In addition, we consider a finite-value quantum annealing, in which the quantum fluctuation remains at the final stage of optimization.
In standard QA, we gradually increase $\rho(t)$ to obtain a single realization among many replicas.
However, as discussed later, a moderate $\rho(t)$ value is beneficial for obtaining improved generalization performance. 
When we do not consider the ``quality'' of the solution, the standard QA is one of the best choices.
The theoretical assurance of the ideal QA toward the optimal solution with the lowest cost function value is well established on the basis of the adiabatic theorem \cite{Suzuki2005}.
However, as in the case of DNN optimization, the quality of the solution is measured using a different scale than the cost function itself, namely the generalization performance.
Therefore, the standard QA method is not necessarily the best choice for optimization of DNNs.
As a result, we inject a finite quantum fluctuation value to attain better generalization performance.

\begin{figure}[t]
\includegraphics[width=0.47\textwidth]{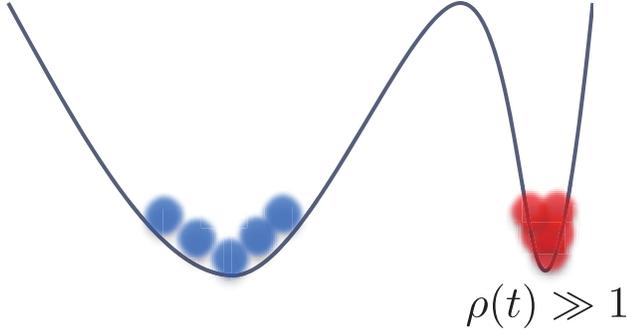}
\caption{Schematic pictures of two local minima and quantum effects.}
\label{figimage2}
\end{figure}

Here, we provide a simple schematic picture for the finite-value QA to attain improved generalization performance.
For simplicity, we assume that a DNN loss function has two local minima:
a sharp local minimum and a wide local minimum.
Both of the depths are the same, as shown in Fig. \ref{figimage2}.
In other words, the first term in the cost function (\ref{RQA}) takes the same values in two local minima.
Let us here consider the favorable solution in the standard QA.
In standard QA, we increase $\rho(t)$ to a very large value.
When the optimization is successfully performed without entrapment in any saddle points or trivial local minima,
we compare the two representative local minima of the cost function (\ref{RQA}).
When most of the realizations of the $M$-replicated DNNs are condensed to the sharp local minimum, the cost function (\ref{RQA}) takes a smaller value compared to the case of the wide local minimum.
Thus, the successful result of the standard QA is absorbed in the sharp local minimum.
In this sense, standard QA is not suitable for optimization of DNNs.
Instead, in finite-value QA, the final value of $\rho(t)$ is set to be finite.
Then, depending on the final value of $\rho(t)$, the resultant solution is allowed to be absorbed into the wider local minimum of the loss function.
In a previous study \cite{Baldassi2016}, $\gamma(t)$ (similar to $\rho(t)$) is referred to as the scoping coefficient and is gradually decreased.

The remaining problem is that, in general, a priori we do not find an adequate strength value for quantum fluctuation.
We propose an adaptive approach for tuning the value of $\rho(t)$ in the next subsection.

\subsection*{Quantum Adam}
We hereafter assume the loss function $L(\mathcal{D}|{\bf w})$ for a training dataset $\mathcal{D}$ as the energy function.
The loss function measures the discrepancy between the ground truth labels ${\bf t}$ and the output ${\bf y}$ predicted by the network.
The gradient of the loss function is computed using the back-propagation method \cite{Rumelhart1986}.
We here employ the stochastic gradient descent method by dividing the training dataset into $M$ minibatches as $\{\mathcal{D}_1,\mathcal{D}_2,\cdots, \mathcal{D}_M\}$.
It is convenient to process a large amount of training data and mitigate the computational cost of the gradient.
We then distribute the minibatch to each Trotter slice $k$.
Following the standard prescription of the Suzuki-Trotter decomposition, we should utilize the same energy function on each Trotter slice.
However, to induce the stochastic ingredients over $M$-replicated DNNs to perform efficient learning, we employ the loss function as $L(\mathcal{D}_k|{\bf w}_k)$ on each Trotter slice $k$.
Thus, we divide the training dataset into $M$ minibatches, where $M$ is the number of Trotter slices.
We then sweep all the minibatches over each Trotter slice in an epoch.
The minibatches are randomly shuffled in each epoch.

We here assume that our procedure is employed in practice in a parallel computing environment.
In the context of the current machine learning environment, parallel computing for learning is sometimes employed for very large datasets.
As in our case, the elastic term $\rho\left\|{\bf w}_k - {\bf w}^*\right\|_2^2$  has been used in parallel computing environments \cite{Zhang2015}.
Another study prepared the master with ${\bf w}$ and updated it by summing over gradients obtained by slaves with ${\bf w}_k$ \cite{Li2015}.

We now address the remaining problem of determining the magnitude of the coefficient $\rho(t)$ of the elastic term.
We exploit the idea of the Adam method, which is often implemented in DNN optimization \cite{Adam2015}, to adaptively change the coefficient.
It accelerates the update when the gradient tends to shrink around the saddle point.
In Adam, instead of the standard gradient descent method (\ref{SGD}), 
\begin{equation}
{\bf w}(t+1) = {\bf w}(t) - \frac{\eta}{\sqrt{\tilde{{\bf v}}(t)}+ \epsilon}\tilde{{\bf m}}(t),
\end{equation}
where $\tilde{{\bf m}}(t) = {\bf m}(t)/(1-\beta_1^t)$, $\tilde{{\bf v}}(t) = {\bf v}_k(t)/(1-\beta_2^t)$,  and
\begin{eqnarray}
{\bf m}(t) &=& (1-\beta_1){\bf m}(t-1) + \beta_1 {\bf g}(t) \\
{\bf v}(t) &=& (1-\beta_2){\bf v}(t-1) + \beta_2 {\bf g}(t)\odot {\bf g}(t).
\end{eqnarray}
Here, ${\bf g}(t)$ is the gradient of the loss function.
The hyperparameters $\beta_1$ and $\beta_2$ are chosen a priori.
The quantity of $\epsilon$ avoids accidental division by zero.
The calculation of the product $\odot$ and the division between vectors are performed in a component-wise manner.
During update iterations, the magnitude of the gradient becomes small around the saddle point.
Then, ${\bf v}(t)$ becomes a vector with small-valued elements.
The coefficient $\eta/\sqrt{\tilde{{\bf v}}(t)}+ \epsilon$ of the effective gradient $\tilde{{\bf m}}(t)$ is then increased.
The updates are then efficiently performed, even around the saddle point.
This is a rough sketch of the learning acceleration provided by Adam.

For tuning $\rho(t)$, we employ a technique similar to one in Adam, in which the coefficient of the effective gradient is adaptively changed as follows:
\begin{equation}
{\bf w}_k(t+1) = {\bf w}_k(t) - \frac{\eta}{\sqrt{\tilde{{\bf v}}_k(t)}+ \epsilon} \tilde{{\bf m}}_k(t)- \frac{\eta \rho}{\sqrt{\tilde{{\bf v}}^q_k(t)}+ \epsilon}\tilde{{\bf m}}^q_k(t),
\end{equation}
where $\tilde{{\bf m}}_k(t)$ and $\tilde{\bf v}_k(t)$ are obtained in the same manner as in Adam, and $\tilde{{\bf m}}^q_k(t) = {\bf m}^q_k(t)/(1-\alpha_1^t)$, $\tilde{{\bf v}}^q_k(t) = {\bf v}^q_k(t)/(1-\alpha_2^t)$ and
\begin{eqnarray}
{\bf m}^q_k(t) &=& (1-\alpha_1){\bf m}^q_k(t-1) + \alpha_1 {\bf g}^q_k(t) \\
{\bf v}^q_k(t) &=& (1-\alpha_2){\bf v}^q_k(t-1) + \alpha_2 {\bf g}^q_k(t)\odot {\bf g}^q_k(t).
\end{eqnarray}
Here, ${\bf g}^q_k(t) = 2{\bf w}_{k}(t)-{\bf w}_{k+1}(t)-{\bf w}_{k-1}(t)$.
Similar to the process followed in Adam, the hyperparameters $\alpha_1$ and $\alpha_2$ are set a priori.
The above update rule adequately tunes the elastic term.
It reads that the coefficient is tuned as $\rho(t) \to \rho/(\sqrt{\tilde{{\bf v}}^q_k(t)}+ \epsilon)$.

Following the standard QA, the weights are randomly initialized in order to search for good candidates for the optimal solution over a relatively wide range.
In other words, in the initial stage of optimization, the weights associated with the different Trotter slices deviate.
Owing to the elastic term, the discrepancies between Trotter slices begin to lessen after several iterations.
In other words, the tunneling effect gradually decays, and the effective coefficient $\rho/(\sqrt{\tilde{{\bf v}}^q_k(t)}+ \epsilon)$ then increases to enhance the tunneling effect again.
Therefore, the above update rule efficiently induces the tunneling effect without directly tuning the value of the mass $\rho$.
We call the above update rule ``quantum Adam'' in the sense that we add the quantum effects stemming from ${\bf g}^q_k(t)$ while tuning the contribution of the effect during the learning.
We emphasize that other gradient methods developed for machine learning, including AdaGrad \cite{Duchi2011}, AdaDelta \cite{Zeiler2012}, RMSprop \cite{RMS2012}, and the Sum of Functions Optimizer \cite{Jascha2014}, can be implemented in conjunction with the quantum effect in the same manner.

In the following section, we demonstrate the effectiveness of quantum Adam by testing it against two datasets: the MNIST handwritten digit dataset \cite{LeCun1998} and the Olivetti face image dataset \cite{Samaria1994}; both are open datasets often used in benchmark tests for machine learning.

\section*{Results}
In this section, we demonstrate the application of quantum Adam to DNNs by using a well-known open dataset.
Although the datasets used in the experiments contain data that are relatively easy to analyze, there are high computational costs incurred when implementing the $M$-replicated DNNs for the realization of quantum Adam.
In this sense, the present study is simply a proof of concept.

For simplicity, we used ReLU as the activation function in the middle layers in all experiments.
We used cross entropy as the cost function for classification and the mean-squared error for auto-encoding in the results shown below.
The weights are initialized with i.i.d. Gaussian samples with a zero mean and deviation $\sqrt{1/N_l}$, where $N_l$ is the number of inputs for each layer $l$.
We use the standard choice of $\alpha_1=\beta_1= 0.9$ and $\alpha_2=\beta_2 = 0.999$.
We set the common initial conditions and performed $M$-independent classical (standard) and quantum Adam tests for comparison.
We then assessed the generalization performance in terms of the average and minimum/maximum of the loss function/accuracy.
\begin{figure}[t]
\includegraphics[width=0.47\textwidth]{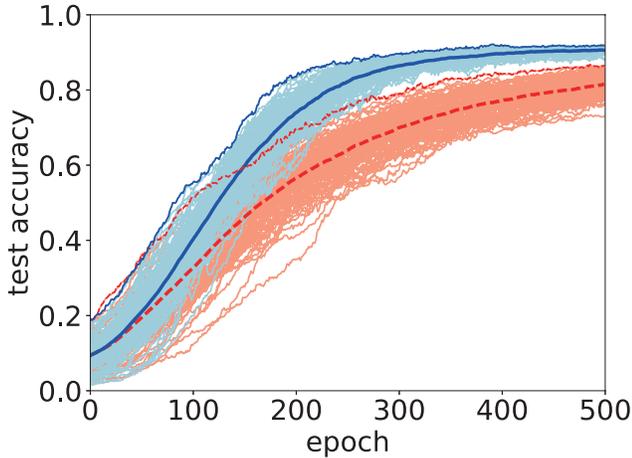}
\caption{Accuracy for test data (red and dashed curves: classical Adam, blue and solid curves: quantum Adam) in single-layer NN for MNIST.
All results from the $M$-replicated systems are indicated by light-colored curves. 
The bold curves denote the average, and the thin curves represent the maximum in the replicated NNs.
The horizontal axis represents the epoch, and the vertical axis represents the accuracy of the test data.}
\label{MNIST}
\end{figure}

The first task was to classify the MNIST $8 \times 8$-pixel images of handwritten digits.
We constructed an all-to-all single-layer neural network (NN) for classifying the handwritten digits.
Figure \ref{MNIST} shows the accuracy with test data for classical and quantum Adam.
We trained the NN by feeding it $500$ data items and setting $M=500$. 
We then measured the accuracy using $1297$ data items.
In this case, we set the coefficient $\rho = 2.0$.
Both the average and the maximum accuracy confirm that quantum Adam is superior to classical Adam.

\begin{figure}[t]
\includegraphics[width=0.50\textwidth]{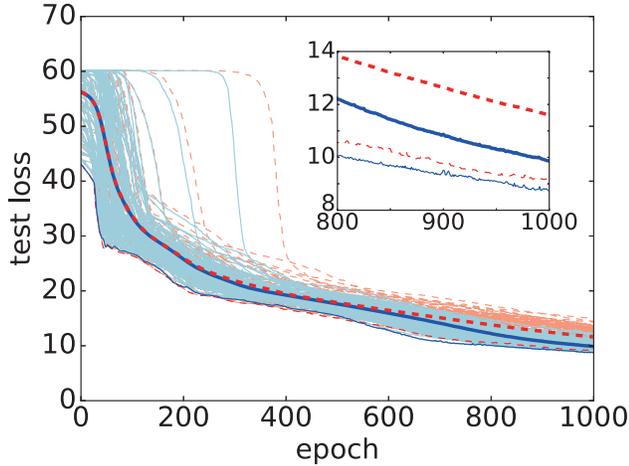}
\caption{Loss function for test data in an auto encoder using MNIST. 
All results from the replicated systems are indicated by light-colored curves. 
The bold and thin curves indicate the average and the minimum in replicated NNs.
The horizontal axis represents the epoch, and the vertical axis represents the loss function of the test data. 
The inset shows an enlarged view of the average loss functions during 800--1000 epochs.}
\label{MNIST_AE}
\end{figure}

The second task was to make the auto encoder. 
It recovers the original input as the output by using MNIST $8 \times 8$-pixel images of handwritten digits.
To encode the handwritten digits, we constructed two-convolution layers with a filter size of three and an output of six channels. 
The middle layer has 96 nodes in this case.
To decode the images, we constructed two deconvolution layers in an inverse manner.
Figure \ref{MNIST_AE} shows the loss function for the test data with classical and quantum Adam.
We trained the NN by feeding it $100$ data items and setting $M=100$. 
We then measured the loss function for $1697$ data items to determine the generalization performance.
In this case, we set the coefficient $\rho = 1.0$.
Both the average and the minimum of the loss function in the replicated systems confirm that quantum Adam is superior to classical Adam.
However, this result might be accidental, as there were no significant improvements in several experiments in terms of the mean-square error.

\begin{figure}[t]
\includegraphics[width=0.50\textwidth]{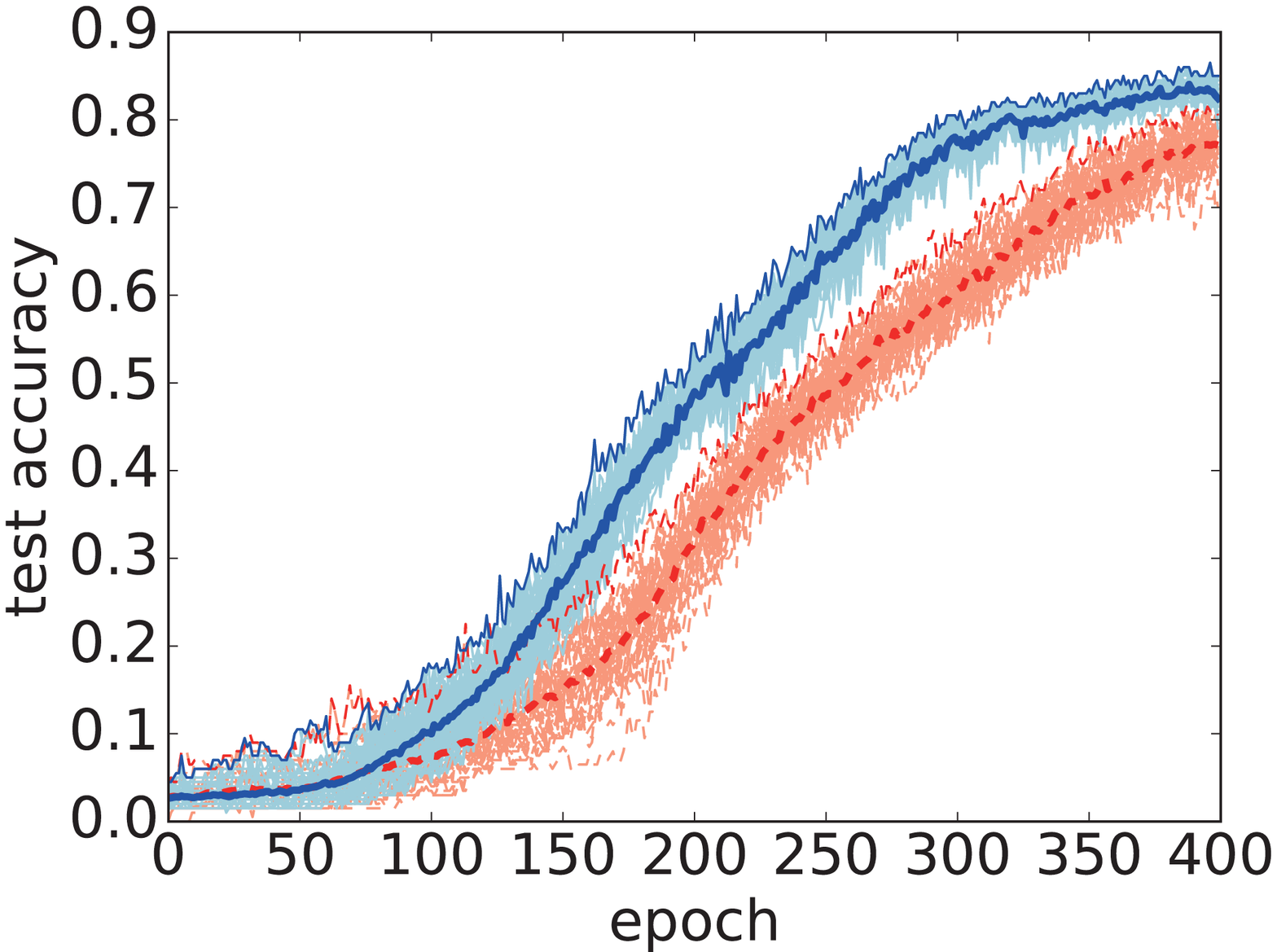}
\caption{Accuracy for test data for classification of Olivetti face images. 
The same curves as those in Fig. \ref{MNIST} are used.
The horizontal axis represents the epoch, and the vertical axis represents the accuracy of the test data.}
\label{Olivetti}
\end{figure}

The third task was to classify the Olivetti $64 \times 64$-pixel images of human faces.
We constructed an all-to-all three-layer ($4096$-$2048$-$1024$-$40$) NN for classifying face images.
Figure \ref{Olivetti} shows the accuracy with the test data for classical and quantum Adam.
We trained the NN by feeding it 200 $data$ points and setting $M=40$. 
We then determined the accuracy using $200$ data items.
In this case, we set the constant $\rho = 1.0$ and performed batch normalization at each layer.
Both the average and the maximum accuracy are evidence that quantum Adam is superior to classical Adam in the last stage of learning.

\section*{Discussion}
We proposed a quantum Adam formulated through a path-integral representation for optimization of DNNs. 
The proposed algorithm generates an elastic term between different realizations of DNNs and could find a better solution in terms of generalization performance than that by classical Adam.
The point is to control the quantum fluctuation by introducing the adaptive change of the coefficient and inducing the wide-flat local minimum by means of the entropy effect, as discussed in the previous studies \cite{Baldassi2016,Baldassi2018}.
In the present study, we directly optimize the $M$-replicated DNNs while dealing with the non-perturbative effect, which allows the quantum tunneling effect.
Although relatively small datasets are used, we demonstrate better generalization performance by considering the optimization with a finite quantum fluctuation strength.
In this sense, our method does not conform to the standard QA method.
The ideal QA might not be the best choice of learning algorithm for DNNs because the resultant solutions are absorbed into a sharp minimum.
In recent development of manufacturing microdevices, QA has been successfully implemented in superconducting qubits, or so-called quantum annealer.
Several experiments have shown that the resultant solutions seem to fall into wide local minima \cite{Johnson2011}.
However, this is due to the freezing phenomena in the quantum annealer, which is a particular problem in the quantum device.
The resultant solutions are closely related to low-energy states with a certain value of quantum fluctuation as pointed out in the literature \cite{Amin2015}.
In other words, the output from the present version of the quantum annealer follows the Gibbs-Boltzmann distribution with a certain value of quantum fluctuations.
In this sense, QA, which is performed in real experiments, can be a choice of learning algorithm.
In addition, the current version of a quantum annealer, the D-Wave 2000Q, implements two optimization techniques by manipulating a certain value of quantum fluctuation, namely quenching, and reverse annealing. 
These two techniques will be available for efficiently attaining better generalization performance in real experiments, as discussed in the literature \cite{Baldassi2018}.

In the present study, we manipulate the optimization in classical computers.
In addition, we select the strength of the quantum fluctuation by employing adaptive change inspired by the Adam method.
The potential performance of quantum Adam emerges in cases with many Trotter numbers that correspond to the number of minibatches.
When we use a small number of minibatches, quantum Adam does not work well.
This is because most of the DNNs fall into the sharp minimizers.
In addition, the $\rho$ value should be tuned adequately.
When we select a $\rho$ value that is too high, the searching range will be narrow, whereas a $\rho$ value that is too small will not lead to a condensed solution. 
We tested three different tasks to assess the performance of quantum Adam in comparison to classical Adam.
The results demonstrate that quantum Adam can provide fairly good performance.
We emphasize that the most important feature of quantum Adam should be its generalization performance.
In machine learning, the purpose of improvements in learning is nothing more than enhancing generalization performance with limited epochs and computational resources.
In quantum Adam, the elastic term aggregates DNNs while learning.
This effect might work to prevent sudden falls into the valley.
In other words, when most of the DNNs are in the wide minimizer, the others do not tend to fall into the sharp minimizer; this can lead to improved generalization performance.

In quantum Adam, we use $M$-replicated DNNs.
In a sense, this seems to be too abundant.
However, when we process a large number of datasets, we distribute each batch to a number of processors or GPUs and establish a consensus to obtain DNNs with high generalization performance.
Our present method is too computationally expensive to implement in the ordinary environments used in a wide range of research efforts, although it might be useful for learning large datasets in parallel computing environments.
In this sense, our algorithm might be helpful even in classical computers.
In future research, we shall test quantum Adam in a parallel computing environment with a large dataset comprising high-dimensional components, and propose another simplified algorithm by elucidating the most significant part of the quantum fluctuations, as in previous studies \cite{Baldassi2016,Baldassi2018}.

We remark on the time complexity of quantum Adam.
The standard assessment of the time complexity of QA can be performed by estimating the energy gap in the time-dependent Hamiltonian.
In our case, through the Suzuki--Trotter decomposition, the problem is reduced to the optimization problem for the cost function with continuous variables.
By considering the rate of convergence to be at a minimum in the feasible set, the classical Adam method has a convergence rate of $O(1/\sqrt{T})$, as shown in the literature \cite{Adam2015}.
We believe that a similar analysis can also be performed for quantum Adam.
In addition, we emphasize that the most important feature of quantum Adam is its generalization performance.
In this sense, the present study triggers a new aspect of QA not for pursuing the minimum of the cost function, but for different optimality measured in a different indicator from the cost function itself. 

Finally, in present study, we demonstrate a potential power of quantum fluctuation, as done by QA.
It promotes ``quality" of solution via optimization with quantum fluctuation.
The standard assessment of the performance of optimization solver is evaluated by the cost function itself.
In particular, the performance of QA has been discussed through the decrease of the cost function.
However, the robustness of the solution can be attained by optimization of the cost function in conjunction with the local entropy as discussed in the literature \cite{Baldassi2016,Baldassi2018}.
The optimization with quantum fluctuation automatically and potentially leads to the robustness of the solution as discussed in the present study.
In the context of machine learning, the generalization performance is robustness of the solution.
In future, deepening the understanding of the quantum fluctuation would promote various approaches in machine learning and beyond.

\bibliography{SR_QMC_ver6}

\section*{Acknowledgements}

The authors would like to thank Shu Tanaka and Muneki Yasuda for many fruitful discussions that contributed to the work.
The present work is financially supported by MEXT KAKENHI Grant No. 15H03699 and 16H04382, and by JST START.

\section*{Author contributions statement}

M.O. conceived and conducted the experiment and analyzed the results.
S. O. tested the previous version of the optimization method, M. T. discussed the possibility of the other applications of our method to industry, S. T. directed the project in our study and investigated the possible design of our method.
All authors discussed the details of the results and reviewed the manuscript. 

\section*{Additional information}
{\bf Competing Interests}: The authors declare that they have no competing interests.
\appendix
\section*{Path integral representation}
By use of the Suzuki-Trotter decomposition, we formulate the path integral representation.
Let us start the following expression of the Suzuki-Trotter decomposition as
\begin{equation}
Z = {\rm Tr}\left\{ \exp\left( - \beta V(\hat{\bf w}) - \frac{\hat{\bf p}^2}{2\rho}\right)\right\} = {\rm Tr}\left\{ \prod_{k=0}^{M-1} \exp\left( - \frac{\beta}{M} V(\hat{\bf w})\right)\exp\left( - \frac{\beta\hat{\bf p}^2}{2\rho M}\right)\right\}.
\end{equation}
We insert the summation over the complete set $\int d{\bf w}_k |{\bf w}_k \rangle \langle {\bf w}_k |$ and $\int d{\bf p}_k |{\bf p}_k \rangle \langle {\bf p}_k |$ where $\hat{\bf w}|{\bf w}_k\rangle = {\bf w}_k |{\bf w}_k\rangle$ and $\hat{\bf p}|{\bf p}_k\rangle = {\bf p}_k |{\bf p}_k\rangle$.
Then we obtain
\begin{eqnarray}
Z=\int d{\bf w}_0 \langle {\bf w}_0 |\int \mathcal{D}{\bf w}\mathcal{D}{\bf p}\prod_{k=1}^{M} \left\{ \exp\left( - \frac{\beta}{M} V(\hat{\bf w}) \right) |{\bf w}_{k}\rangle \langle {\bf w}_{k}|\exp\left(- \frac{\beta\hat{\bf p}^2}{2\rho M}\right) |{\bf p}_{k}\rangle \langle {\bf p}_{k}| \right\}|{\bf w}_0\rangle
\end{eqnarray}
This expression can be reduced to
\begin{eqnarray}
Z
\propto
\int d{\bf w}_0 \int \mathcal{D}{\bf w}\mathcal{D}{\bf p} \prod_{k=1}^{M}\left\{  \exp\left( - \frac{\beta}{M} V({\bf w}_{k}) \right) \exp\left( i{\bf p}_{k}({\bf w}_{k}-{\bf w}_{k-1})\right) \exp\left(- \frac{\beta{\bf p}_{k}^2}{2\rho M}\right)  \right\} 
\end{eqnarray}
where we have used
\begin{equation}
\langle {\bf w}_{k'} | {\bf p}_{k} \rangle = \exp\left( i {\bf p}_k {\bf w}_{k'} \right).
\end{equation}
Manipulation of the Gaussian integral with respect to ${\bf p}_k$ yields
\begin{eqnarray}
Z \propto
 \int d{\bf w}_0 \int \mathcal{D}{\bf w} \prod_{k=1}^{M} \exp\left( - \frac{\beta}{M} V({\bf w}) -  \frac{M \rho}{2\beta}\left\| {\bf w}_k -{\bf w}_{k-1}\right\|_2^2 \right).
\end{eqnarray}

\section*{Strong limit of $\rho(t)$}
First we consider the Fourier transformation on the discrepancy from the center of weights ${\bf w}^*$ as
\begin{equation}
{\bf w}_k = {\bf w}^* + \frac{1}{\sqrt{M}}\sum_{r=0}^{M-1} {\bf a}_r {\rm e}^{i2\pi kr/M},
\end{equation}
where ${\bf a}_r = {\bf a}_{M-r}$ because ${\bf w}_k$ is a real vector.
Then the elastic term is diagonalized as 
\begin{equation}
\left\| {\bf w}_k - {\bf w}_{k-1}\right\|_2^2 = 2 \sum_{r=1}^{[M/2]} {\bf a}_r{\bf a}_{M-r} \left(1-\cos\left(\frac{2\pi r}{M}\right)\right).
\end{equation}
where we have used $\sum_{k=0}^{M-1}{\rm e}^{i2\pi kr/M} = M\delta(r)$.
When $\rho(t) \gg 1$,  the exponentiated elastic term is reduced to
\begin{equation}
\prod_{k=1}^M \exp\left( - \frac{ M \rho(t)}{2\beta}\left\| {\bf w}_k - {\bf w}_{k-1}\right\|_2^2\right) = \prod_{r=1}^{[M/2]} \exp\left( - \frac{M \rho}{\beta}{\bf a}_r{\bf a}_{M-r} \left(1-\cos\left(\frac{2\pi r}{M}\right)\right)\right).
\end{equation}
We find that ${\bf a}_r$ follows the Gaussian distribution.
We then perform the inverse Fourier transformation and attain 
\begin{equation}
\prod_{k=1}^M \exp\left( - \frac{ M \rho(t)}{2\beta}\left\| {\bf w}_k - {\bf w}_{k-1}\right\|_2^2\right) = \prod_{k=1}^{M} \exp\left( - \frac{\beta}{2}\sum_{k,k'}({\bf w}_{k}-{\bf w})V_{k',k'}({\bf w}_{k'}-{\bf w})\right).
\end{equation}
In $M \to \infty$, we use $2\pi r/M = x$ and $2\pi/M = dx$
\begin{equation}
\frac{1}{\beta}V^{-1}_{kk'} = \sum_{r}\frac{\beta}{2M \rho\left(1-\cos\left(\frac{2\pi r}{M}\right)\right)} {\rm e}^{i2\pi (k-k')r/M}  = \frac{\beta}{2 \rho}\int^{2\pi}_{0} \frac{dx}{2\pi}\frac{{\rm e}^{i(k-k')x}}{1-\cos x} .
\end{equation}

\end{document}